%% file: Paper.tex
\documentclass[runningheads]{llncs}
\usepackage{graphicx}

\usepackage{mathtools}
\usepackage{multirow}
\usepackage{tikz}
\newcommand{\eg}{\emph{e.g} }
\newcommand{\ie}{\emph{i.e} }
\usepackage{amssymb}
\usepackage{booktabs}
\usepackage{caption}
\usepackage{bbm}
\usepackage{subcaption}
\usepackage{array}

\usepackage{marvosym}

\definecolor{cvprblue}{rgb}{0.21,0.49,0.74}
\usepackage[breaklinks,colorlinks,citecolor=cvprblue]{hyperref}
\usepackage{cleveref}

\definecolor{negative}{HTML}{436B8E}
\definecolor{positive}{HTML}{6AB2A4}

\newcommand*\circled[1]{\tikz[baseline=(char.base)]{
            \node[shape=circle,fill,inner sep=1pt] (char) {\textcolor{white}{#1}};}}

\begin{document}
\title{FedMedICL: Towards Holistic Evaluation of Distribution Shifts in Federated Medical Imaging}

\titlerunning{FedMedICL: Holistic Evaluation in Federated Medical Imaging}

%
\author{Kumail Alhamoud\inst{1}\thanks{Equal contribution. Yasir did part of this work while interning at Oxford.} \and Yasir Ghunaim\inst{2,3}$^\star$ \and Motasem Alfarra\inst{2} \and Thomas Hartvigsen\inst{4} \and Philip Torr\inst{3} \and Bernard Ghanem\inst{2} \and Adel Bibi\inst{3} \and Marzyeh Ghassemi\inst{1}}

%
\authorrunning{Kumail Alhamoud \emph{et al}.}
%
%
\institute{MIT, USA \and King Abdullah University of Science and Technology (KAUST), Saudi Arabia \and University of Oxford, UK \and University of Virginia, USA \\ Correspondence to: \email{kumail@mit.edu, yasir.ghunaim@kaust.edu.sa} }
\maketitle              

\input{LatexSource/sections/0_abstract}
\input{LatexSource/sections/1_introduction}
\input{LatexSource/sections/2_related_work}
\input{LatexSource/sections/3_fedmedicl}
\input{LatexSource/sections/4_results}
\input{LatexSource/sections/5_pandemic_simulation}
\input{LatexSource/sections/6_conclusions}
\begin{credits}
\subsubsection{\ackname}
The authors thank Haoran Zhang for his feedback. Kumail was supported by a Jameel Clinic Fellowship and a Saudi Arabian Cultural Mission Scholarship. Yasir was supported by Saudi Aramco. Marzyeh is supported in part by an NSF CAREER award, a CIFAR AI Chair held at the Vector Institute, and the Gordon and Betty Moore Foundation. This project was also supported by the KAUST Office of Sponsored Research (OSR) under Award No. OSR-CRG2021-4648 and a UKRI grant Turing AI Fellowship (EP/W002981/1). Philip thanks the Royal Academy of Engineering for their support.

\subsubsection{\discintname}
The authors have no competing interests to declare that are relevant to the content of this article.
\end{credits}



\bibliographystyle{splncs04}
\bibliography{Paper}

\input{SupplementaryFile/appendix}

\end{document}

%% file: LatexSource/sections/0_abstract.tex
\begin{abstract}
For medical imaging AI models to be clinically impactful, they must generalize. However, this goal is hindered by \emph{(i)} diverse types of distribution shifts, such as temporal, demographic, and label shifts, and \emph{(ii)} limited diversity in datasets that are siloed within single medical institutions.
While these limitations have spurred interest in federated learning, current evaluation benchmarks fail to evaluate different shifts simultaneously.
However, in real healthcare settings, multiple types of shifts co-exist, yet their impact on medical imaging performance remains unstudied.
In response, we introduce FedMedICL, a unified framework and benchmark to holistically evaluate federated medical imaging challenges, simultaneously capturing label, demographic, and temporal distribution shifts. We comprehensively evaluate several popular methods on six diverse medical imaging datasets (totaling 550 GPU hours). Furthermore, we use FedMedICL to simulate COVID-19 propagation across hospitals and evaluate whether methods can adapt to pandemic changes in disease prevalence. We find that a simple batch balancing technique surpasses advanced methods in average performance across FedMedICL experiments. This finding questions the applicability of results from previous, narrow benchmarks in real-world medical settings. 
Code is available at: \url{https://github.com/m1k2zoo/FedMedICL}.

\keywords{Federated Learning  \and Continual Learning \and Distribution Shifts.}
\end{abstract}

%% file: LatexSource/sections/1_introduction.tex
\input{LatexSource/figures/setup_and_construction}

\section{Introduction}
\label{sec:intro}
Machine learning increasingly impacts medical imaging~\cite{kamnitsas2017efficient,yala2019deep}. Yet, most FDA-approved models are validated on datasets unrepresentative of real demographic distributions, leading to potential inaccuracies in medical diagnostics~\cite{ebrahimian2022fda}. This lack of diversity in data is problematic when models encounter distribution shifts~\cite{yang2023change}. These shifts reflect discrepancies in data characteristics between the training/validation datasets and actual clinical settings. Thus, they can degrade the performance of diagnostic models across diverse patient populations~\cite{pooch2020can,tschandl2018ham10000}.

Merging medical datasets from different hospitals, known as data pooling, can enhance training data diversity and address distribution shifts~\cite{petkova2020pooling}. 
However, the practical implementation of this approach is often challenged by institutional policies, and slow data access processes~\cite{ogier2022flamby,white2022data}, which severely limit data sharing across healthcare centers.
As a result, there is a heavy reliance on smaller, \emph{siloed} datasets, that hinder the deployment of generalizable medical imaging models~\cite{ebrahimian2022fda}. 

As exemplified in \Cref{subfig:FedMedICL_pull_figure}, we investigate two key factors impacting the development of robust medical imaging models: the challenges posed by distribution shifts and the siloed nature of medical data. 
We frame the complications that arise from this complex setup under three types of shifts. These types are:

\vspace{3pt}\noindent\textbf{\circled{1} Label Imbalance}: 
In individual institutions, some medical conditions are more prevalent, leading to models failing on rarer conditions \cite{alshammari2022long,marrakchi2021fighting}. For instance, a skin disease classification model trained in one region may perform poorly on conditions more common in another geographic region~\cite{baghestani2005skin,paek2012skin}.

\vspace{3pt}\noindent\textbf{\circled{2} Demographic Imbalance}: 
Patient demographics vary significantly among healthcare centers. For example, hospitals in American suburbs, which have a higher population of seniors~\cite{pekmezaris2013aging}, may have fewer data samples from younger patients, leading to models that are less effective for younger demographics~\cite{zong2023medfair}.

\vspace{3pt}\noindent\textbf{\circled{3} Temporal Distribution Shifts}:
Medical data is also affected by temporal changes, including shifts in disease prevalence and patient demographics~\cite{vokinger2021continual}. 
An example is the emergence of COVID-19, requiring models to remain effective amidst these shifts.
Given these challenges, one cannot help but question:
\begin{center}
    \emph{Are current training and evaluation protocols truly equipped to ensure that deployed medical imaging models perform robustly across different institutions?}
\end{center}
Federated learning is one proposed solution, enabling collaborative training across disparate datasets~\cite{ogier2022flamby}, which could improve robustness while maintaining data separation. Yet, current evaluations of federated learning do not explicitly consider label, demographic, and temporal shifts simultaneously. Addressing this gap, we contribute a novel problem setup designed to simulate versatile medical scenarios. We also provide comprehensive benchmarking of existing methods, and we use our setup to emulate federated learning under pandemic conditions.

\vspace{3pt}\textbf{Problem and 
Benchmark Contribution.} We introduce FedMedICL (Federated Medical Imaging with Continual Learning), a dual-function problem formulation. ({\textbf{i}}) It offers a unified approach for modeling label, demographic, and temporal shifts in medical imaging. ({\textbf{ii}}) It acts as a comprehensive testbed for evaluating the efficacy of federated learning in diverse healthcare settings. FedMedICL uses demographic metadata to automatically create federated and continual learning tasks, resulting in a realistic benchmark. We aim to foster reproducible research in this vital setup by releasing our extendable code.

\vspace{3pt}\textbf{Pandemic Spread Simulation Contribution.} 
We showcase a practical application of FedMedICL: evaluation of model adaptation to a novel disease under pandemic conditions. In our setup, different hospitals may have access to COVID-19 datasets that grow at different rates over time. This evaluation is unexplored in prior federated learning research.

%% file: LatexSource/figures/setup_and_construction.tex
\begin{figure}[t]
    \centering
    \begin{subfigure}[b]{0.48\textwidth}
        \centering
        \includegraphics[width=\textwidth]{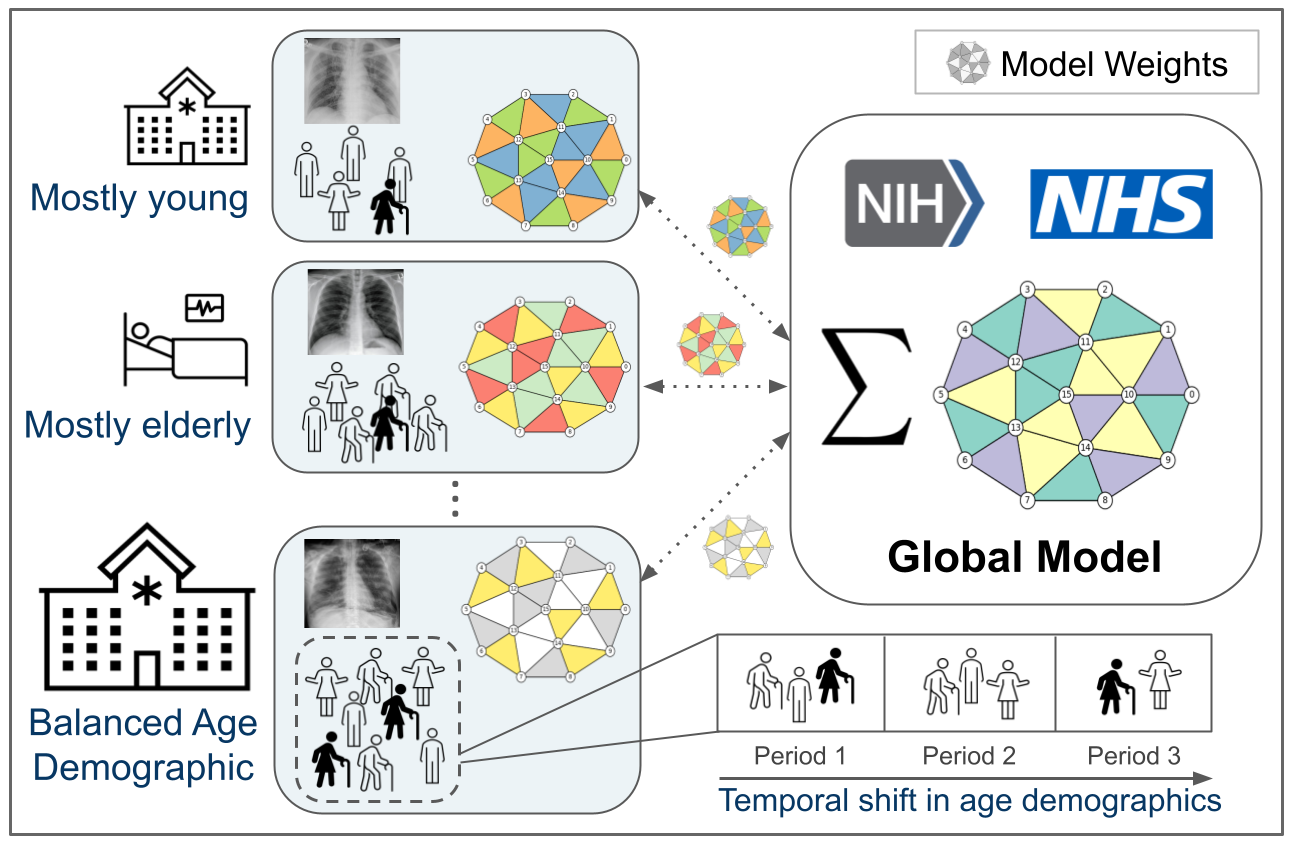}
        \caption{} 
        \label{subfig:FedMedICL_pull_figure}
    \end{subfigure}
    \hfill 
    \begin{subfigure}[b]{0.48\textwidth}
        \centering
        \includegraphics[width=\textwidth]{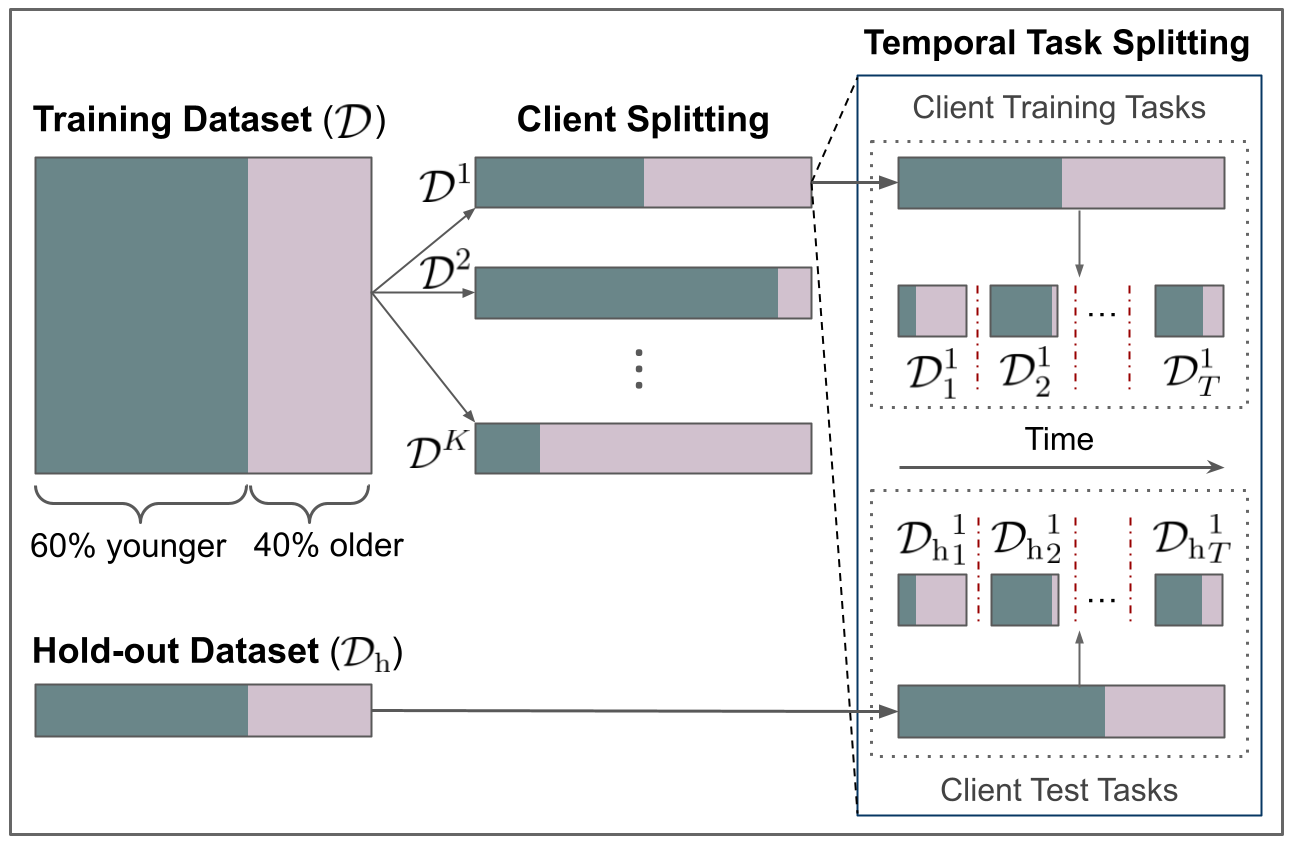}
        \caption{}
        \label{subfig:FedMedICL_setup}
    \end{subfigure}
    \caption{(a) \textbf{Problem Setup.} We model a federated medical imaging scenario, in which siloed hospitals experience demographic imbalances and temporal shifts. (b) \textbf{FedMedICL Benchmark Construction.} We construct client datasets ($\mathcal{D}^1$ to $\mathcal{D}^K$), each representing a hospital with unique demographic characteristics and temporal training tasks ($\mathcal{D}^i_1$ to $\mathcal{D}^i_T$). We evaluate models on temporally aligned test tasks for testing adaptability to local demographic shifts, and on a hold-out set ($\mathcal{D}_h$) to evaluate generalization to diverse demographics.}
    \label{fig:combined_fedmedicl}
\end{figure}

%% file: LatexSource/sections/2_related_work.tex
\section{Related Work}\label{sec:related_work}
\textbf{Distribution Shifts in Medical Imaging.} Research in long-tailed recongition~\cite{alshammari2022long}, fairness~\cite{Sagawa*2020Distributionally}, and continual learning~\cite{chaudhry2019tiny} targets label imbalance~\cite{marrakchi2021fighting}, demographic~\cite{yang2023change}, and temporal shifts~\cite{derakhshani2022lifelonger} respectively, yet typically operates in isolation. Recent works like SubpopBench~\cite{yang2023change} have made advances by formalizing subpopulation shifts, yet they do not study them together, and do not model temporal shifts. Similarly, MEDFAIR~\cite{zong2023medfair} comprehensively evaluates demographic fairness but leaves label imbalance and temporal shifts unexplored. Furthermore, these studies typically overlook the siloed nature of medical data, a crucial aspect of real-world healthcare settings. In contrast, we offer a holistic approach to evaluate label, demographic, and temporal distribution shifts in medical imaging, while also considering the siloed nature of medical data.

%% file: LatexSource/sections/3_fedmedicl.tex
\section{FedMedICL: Problem and Benchmark}
\label{sec:mathematical_formalization}

We study three types of distribution shifts associated with classification for real-world federated medical imaging. Individual hospitals possess small, isolated datasets, leading to label and demographic imbalances (types \circled{1} and \circled{2}). Medical data continually evolves, necessitating models that adapt to temporal shifts (type \circled{3}). To replicate these real-world conditions, our benchmark consists of two components that follow the federated learning and continual learning literature~\cite{mcmahan2017communication,derakhshani2022lifelonger}: Client Splitting, to simulate the distribution of data across medical institutions, and Temporal Task Splitting, to model the evolution of data over time. 
Next, we formalize the setup for federated and continual learning.

\subsection{Background on Federated and Continual Learning}
Assume we are given an input image $x\in \mathcal X$ and its ground truth diagnosis label $y\in \mathcal Y$ with $\mathcal Y=\{1, 2, \dots, L\}$. We aim to build a classifier $f_\theta: \mathcal X \rightarrow \mathcal P(\mathcal Y)$, parameterized by $\theta$, that can correctly classify $x$, where $\mathcal P(\mathcal Y)$ is the probability simplex over the label space.
Furthermore, medical data is often coupled with a set of attributes $a_1, a_2, \dots a_m \in \mathcal A$, \eg, the age or sex of patients. 

\vspace{3pt}\textbf{In federated learning,} client splitting divides a dataset $\mathcal D$ into $K$ segments $\{\mathcal{D}^1, \dots \mathcal{D}^K\}$, each corresponding to a hospital. 
Each hospital may have a different distribution of attributes.
Let $\mathcal D^k$ be the training distribution at hospital $k$. $\mathcal D^k$ is characterized by the set of probabilities of observing the different attributes at that hospital, denoted by $\{p_i^k\}_{i=1}^m$ where $\mathbb P_{(x, y)\sim\mathcal D^k}(a=a_i) = p_i^k$.
That is, $p_i^k\geq0$ is the probability that the attribute $a_i$ is observed at hospital $k$.

\vspace{3pt}\textbf{In continual learning,} temporal task splitting divides a hospital's dataset $\mathcal{D}^i$ into $T$ sequential segments, $\{\mathcal{D}^i_1, \dots \mathcal{D}^i_T\}$, which are called temporal tasks. 

\subsection{Benchmark Construction Methodology}
We propose a benchmark construction method that precisely mirrors the complex scenarios in real-world medical federated and continual learning tasks. 

\vspace{3pt}\textbf{Client Splitting.}
We introduce a scalable method to simulate the distribution of medical data across \(K\) clients, each representing a medical institution, by dividing them into two types: \textit{Balanced} and \textit{Skewed}. A Balanced client \( k \) maintains a homogeneous demographic distribution, defined as \( \forall i, j \in \{1, \ldots, m\}, \, p_i^k \approx p_j^k \). Conversely, a Skewed client \( k \) is marked by an uneven distribution, with at least one attribute \( a_i \) having a significantly higher than average probability, \ie, \( p_i^k \gg 1/m \). We create a mix of Balanced and Skewed clients to reflect the diverse demographic imbalances both within and across different hospitals, mimicking the heterogeneity of real-world medical data.

\vspace{3pt}\textbf{Temporal Task Splitting.} 
We now address the temporal aspect of medical data, necessitating continual learning~\cite{derakhshani2022lifelonger}. We define the \textit{localized split}, which models demographic-based temporal shifts. We also consider novel disease emergence, which is a label-based temporal shift.

\noindent \textit{1- Localized Split.} 
Hospital admission rates for different demographic groups vary seasonally, potentially due to local factors. 
To model these localized factors, we let $p_i^k \neq p_i^{l} \,\, \forall l \neq k$. That is, the distribution of attributes at each client evolve independently and differently from other clients, resulting in cross-institutional data imbalances.
Examples of healthcare scenarios that can be modeled by our localized split include demographic variations in emergency rooms \cite{chaaban2017demographic}. 

\noindent \textit{2- Novel Disease Split.}\label{novel_split}
We aim to simulate the label imbalance associated with the emergence of a new disease, in which different geographic regions experience different rates of disease spread~\cite{sun2020impacts}. Given a novel disease label \( y_{\text{new}} \), we define $T$ tasks characterized by $\{\mathcal{D}^i_1, \dots \mathcal{D}^i_T\}$ for each client's dataset $\mathcal{D}^i$. 
During $T=1$, we have \(\forall i, \mathbb{P}_{(x, y)\sim\mathcal D^i_1}(y = y_{\text{new}}) = 0 \)
for all clients, while in the following tasks
some clients encounter a distribution shift with $\exists i, j \,\, \text{s.t.}$ \(\mathbb{P}_{(x, y)\sim\mathcal D^i_{j > 1}}(y = y_{\text{new}}) > 0  \), simulating the introduction of a novel disease into the cross-institutional dataset. 

\subsection{Evaluation and Datasets}

\vspace{3pt}\textbf{Baselines.} We implement baseline methods from various fields: data augmentation (MixUp~\cite{zhang2018mixup}), domain generalization (SWAD~\cite{cha2021swad}), continual learning (ER~\cite{chaudhry2019tiny}), group-imbalanced learning (GB for group-balancing), and class-imbalanced learning (CB for class-balancing, CRT~\cite{kang2019decoupling}).
Each method tackles a single challenge and is unable to handle all scenarios that FedMedICL simulates. Therefore, we adapt these methods by augmenting them with federated averaging~(FedAvg)~\cite{mcmahan2017communication}. The modified version of each method is prefixed with ``F-" (\eg F-CRT) throughout the paper.
For comprehensive comparisons, we implement local training with Empirical Risk Minimzation~(ERM), which does not use federated learning, serving as a reference point for minimum expected performance. Following MEDFAIR~\cite{zong2023medfair}, we used a ResNet-18 backbone for all the baselines.

\vspace{3pt}\textbf{Datasets.}
We utilize five public medical imaging datasets: CheXpert~\cite{irvin2019chexpert}, Fitzpatrick17k~\cite{groh2021evaluating}, HAM10000~\cite{tschandl2018ham10000}, OL3I~\cite{zambrano2021opportunistic}, and PAPILA~\cite{kovalyk2022papila}. 
We also construct a new dataset, CheXCOVID, combining images from CheXpert and the COVID-19 image data collection~\cite{cohen2020covid}. 
Our choice of datasets is based on their scale, coverage of a wide range of imaging tasks, and availability of demographic metadata. A table summarizing these datasets is included in the Appendix. 

\vspace{3pt}\textbf{Reported Metric.}
These datasets exhibit skewed class distributions, so a naive classifier that overpredicts the majority label achieves a high overall accuracy. Thus, we follow the Long-Tailed Recognition (LTR) literature and measure
the LTR Accuracy~\cite{alshammari2022long}:
\noindent$
    \text{LTR} = \frac{1}{L}\sum_{j=1}^{L}\frac{1}{N_j}\sum_{\forall i \,\text{s.t.}\, y_i=j}\mathbbm{1}\{\text{arg}\max_j f^j_\theta(x_i) = j\},
$
where $N_j$ denotes the number of instances in the $j^{th}$ class. To quantify label imbalance, we report the Imbalance Factor (IF), $\text{IF} = \frac{\max_i N_i}{\min_i N_i}$~\cite{alshammari2022long}, in \Cref{fig:localized_experiment}.

\vspace{3pt}\textbf{Evaluation.} 
As shown in \Cref{subfig:FedMedICL_setup}, we evaluate adaptation to local shifts via temporally aligned test tasks, and generalization across diverse demographics via a hold-out set $\mathcal{D}_h$ that preserves the original dataset's demographic distribution.

%% file: LatexSource/sections/4_results.tex
\section{Benchmark Results}
\label{sec:benchmark_experiments}
We evaluate the baselines using our localized split framework, detailed in Section~\ref{sec:mathematical_formalization}. This framework models distributed hospitals experiencing independent demographic shifts over time. In this section,  demographic groups are defined by skin type for PAPILA and age brackets (\eg 0-40 years) for other datasets, but our framework supports alternative attributes such as sex.
 \input{LatexSource/figures/test_curves_localized}

\vspace{3pt}\textbf{Setup.} 
We set up $K=10$ clients per dataset, except the largest dataset CheXpert with $K=50$. This choice ensures similar dataset sizes across clients, providing a fair comparison across datasets. Employing our localized split strategy, we assign $T=4$ sequential training tasks to each client to simulate seasonal changes in disease patterns and hospital admissions. For example, the prevalence of seasonal flu can alter the demographic composition of hospital patients during flu season. Transitioning to a new task represents evolution in patient age demographics, except for PAPILA, where skin demographics are used. Each client receives $T=4$ local test sets matching their training data demographics, as depicted in \Cref{subfig:FedMedICL_setup}. During each task, training includes multiple communication rounds, each with $M=5$ iterations and a batch size of 10, followed by federated averaging. At the end of every training task, we assess each client’s model on all previously encountered tasks, following Equation (2) in \cite{diaz2018don}, and report the mean LTR across clients in \Cref{fig:localized_experiment}. After the final task, we evaluate each client's final model on a shared test set $\mathcal D_h$, representing a global demographic distribution. This evaluation is reported in \Cref{fig:ood_group_incremental}. 

\vspace{3pt}\textbf{Results.} 
\Cref{fig:localized_experiment} compares methods across multiple datasets and diverse demographic distributions. Our evaluation under the FedMedICL setup reveals that the straightforward class-balancing approach (F-CB, shown in green) surpasses all other methods in 5/6 datasets. Notably, advanced algorithms such as F-SWAD and F-CRT, despite their strong performance in previous benchmarks like SubpopBench~\cite{yang2023change} and MEDFAIR~\cite{zong2023medfair}, fall short in every examined dataset. This discrepancy supports our hypothesis: progress in tackling individual types of shift is not sufficient for deployment in federated settings with simultaneous distribution shifts. Specifically, SWAD's dense stochastic weight averaging requires many iterations to converge, making it suboptimal for FedMedICL scenarios where quickly adapting to new distributions is crucial. Similarly, CRT's two-staged paradigm is cumbersome for the dynamic nature of federated and continual learning, which require frequent client communication and swift adaptation. Results in \Cref{fig:ood_group_incremental}, examining the final model at $T=4$ against a hold-out set, align with these findings. It suggests that better performance on local client data correlates with better out-of-distribution demographic generalization.
\input{LatexSource/figures/holdout_bars_localized}

\textbf{Disparities across Datasets.} We explore why performance varies across FedMedICL datasets. The class-balancing method (F-CB, shown in green) is the most effective in most datasets except PAPILA, where the group-balancing method (F-GB, shown in orange) excels. We hypothesize that this variance relates to the different types of imbalance present in each dataset: group imbalance in PAPILA and class imbalance in other datasets. For example, the HAM10000 dataset has $7$ labels with an imbalance factor (IF) of $58.3$, which indicates a severely skewed label distribution. Similarly, CheXpert has only $2$ labels with IF $=9.6$, meaning that one label encompasses around $90\%$ of the dataset. This is why class-balancing is effective in these datasets. An interesting case is OL3I, having $2$ labels with IF $=22.1$, which is more skewed than CheXpert. Due to its extreme imbalance, methods including F-CB achieve around $50\%$ LTR, which degenerates to the case of simply predicting the majority class. 

In PAPILA, we hypothesize that the variation in image characteristics related to age is more significant than class imbalance. We group images by age and compute the contrast feature, which measures the variation in gray-level intensity between neighboring pixels~\cite{haralick1973textural}.
We compute the Cohen's D value~\cite{cohen2013statistical} to compare image contrast features between older adults (60+) and younger adults (below 60). The values are 0.01, 0.14, and 0.40 for CheXpert, HAM10000, and PAPILA respectively. The higher value for PAPILA confirms the significance of differences between the two age groups in this eye fundus dataset. Our analysis aligns with ophthalmological findings that the eye's structure varies with age~\cite{cavallotti2008age}. Finally, previous benchmarks found a high accuracy gap between age-based groups in PAPILA~\cite{zong2023medfair}, further confirming the age variation challenge in its fundus images. 

%% file: LatexSource/figures/test_curves_localized.tex
\begin{figure*}[t] 
  \centering
  \includegraphics[width=\textwidth]{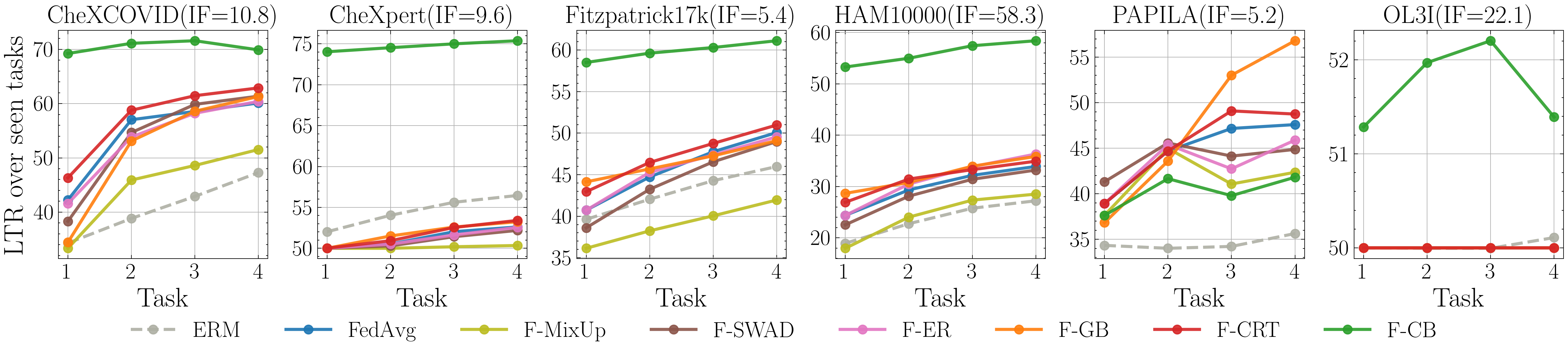}
  \caption{\textbf{Benchmarking Demographic Shifts.} 
We simulate age-based demographic changes over time and benchmark methods in different datasets. The mean LTR accuracy across clients is reported for each method. Except on the PAPILA dataset, no method reliably competes with the simple F-CB baseline.}
  \label{fig:localized_experiment}
\end{figure*}

%% file: LatexSource/figures/holdout_bars_localized.tex
\begin{figure*}[t]   
    \centering
    \includegraphics[width=\textwidth]{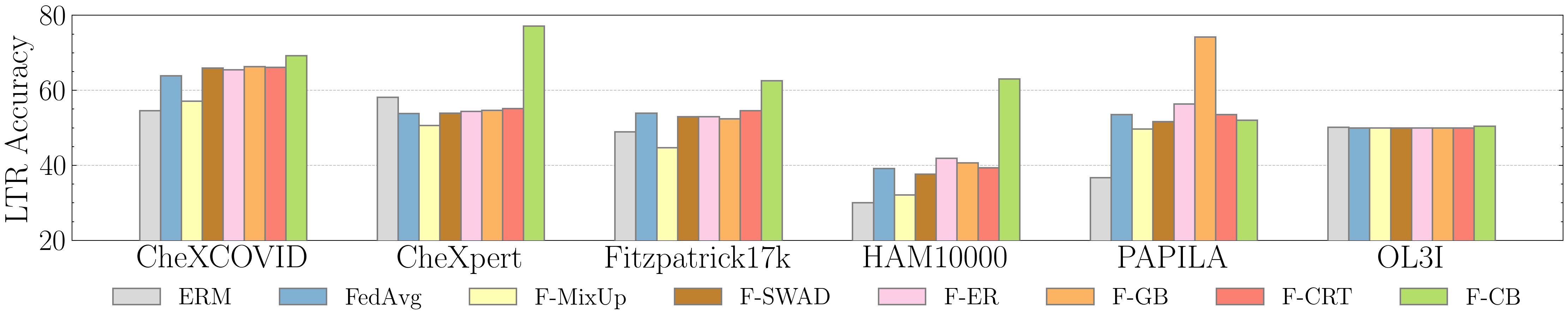}
    \caption{\textbf{Performance on New Demographic Distributions.} 
    We report the LTR accuracy on a hold-out test set, as shown in \Cref{subfig:FedMedICL_setup}. Results are averaged across clients. No method consistently generalizes better than F-CB. 
    }
      \label{fig:ood_group_incremental}
\end{figure*}

%% file: LatexSource/sections/5_pandemic_simulation.tex
\input{LatexSource/figures/novel_disease}
\section{Spread of the COVID-19 Pandemic Between Hospitals}
\label{sec:covid}
FedMedICL can simulate novel disease outbreaks and evaluate algorithms under such conditions. We simulate COVID-19 spreading between hospitals using a subset of our proposed CheXCOVID dataset, which augments CheXpert with COVID-19 samples. Following the Novel Disease Split, introduced in Section~\ref{sec:mathematical_formalization}, we consider a scenario with $K=5$ hospitals. Initially, X-ray screenings across hospitals do not include COVID-19 samples. Over time, the number of samples grows for each hospital, but the rate of increase in COVID-19 admissions varies by geographic region, and hence by hospital~\cite{sun2020impacts}. More formally, we assign $T=4$ sequential training tasks to each hospital, mimicking the evolution of the pandemic. For each task, the number of COVID cases grows. We assume Task 1 is pre-COVID: no hospital has any COVID cases in their training set. Three hospitals experience rapid surges, where approximately $0\%$, $10\%$, $50\%$, and $90\%$ of patients have COVID for each step, respectively. Meanwhile, the remaining two hospitals have slower growth rates, with $0\%$ COVID for the first two tasks, then 10\% and 50\% for the final two. All models are tested on COVID and non-COVID samples at the end of every task. The pivotal question we address is: \emph{How can our simulated hospitals, experiencing varying rates of COVID-19 admission, utilize federated learning to enhance the detection of the novel disease?} 

\textbf{Results.}
\Cref{fig:novel_disease} presents our findings on Non-COVID labels (\ie accuracy on CheXpert labels) and COVID label (recognition of COVID-19). After the pre-COVID training stage (Task 1), all hospital models fail to recognize the novel class. As the tasks progress, ERM exhibits severe forgetting of the Non-COVID conditions, indicated by the severe drop in the 'Non-COVID Performance' graph, and demonstrates slow adaptation to COVID-19, as shown by the limited improvement in Task 2 on the 'COVID Performance' graph. F-ER maintains stable performance on non-COVID labels, which aligns with expectations, since F-ER is designed to mitigate forgetting. However, it has a relatively flat curve for COVID detection in the first two tasks, suggesting poor performance in early pandemic stages. Conversely, F-CB and F-GB quickly adapt to COVID-19 but at the expense of forgetting previously known diseases. 
Hence, there is a need for new methods that can balance high plasticity, allowing for the swift detection of emerging pathogens like COVID-19, with sufficient stability, which maintains performance on established conditions.

%% file: LatexSource/figures/novel_disease.tex
\begin{figure*}[t]   
    \centering
    \includegraphics[width=\textwidth]{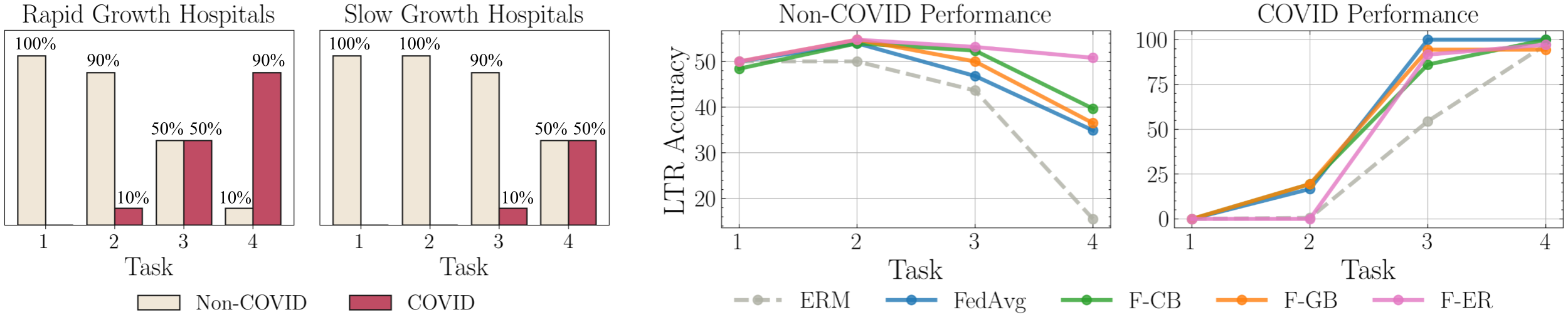}
    \caption{\textbf{Adaptation under Pandemic Conditions.} 
    We simulate two types of hospitals experiencing COVID-19 emergence over four tasks. We report performance on non-COVID labels, in addition to examining how various methods perform in recognizing the novel COVID-19 disease across the four time steps.  
    }
\label{fig:novel_disease}
\end{figure*}

%% file: LatexSource/sections/6_conclusions.tex
\section{Conclusions and Future Research}
\label{sec:conclusion}
We envision FedMedICL as a first step towards holistic evaluation of federated medical imaging. It paves the way for research on distribution shifts in siloed medical datasets, showing that a simple batch-balancing technique outperforms existing methods. Our experiments focus on variations in age and skin type, motivated by evidence of significant performance disparities across these attributes~\cite{zong2023medfair}. Yet, we have designed FedMedICL with flexibility in mind, enabling easy integration of additional singular attributes (e.g., device manufacturer) and dual attributes (e.g., intersection of sex and age). FedMedICL can also easily extend to any modality, such as text and tabular data.  We encourage researchers to expand our benchmark to include more data attributes and  modalities.

%% file: SupplementaryFile/appendix.tex
\newpage
\section*{Appendix}

\begin{table}[h]
\caption{\textbf{Summary of Datasets and Selected Metadata}}
\centering
\setlength{\tabcolsep}{10pt}
\begin{tabular}{@{}lccc@{}} 
\toprule
\textbf{Dataset} & \textbf{\# Labels} & \textbf{Metadata} & \textbf{Imbalance Factor (IF)} \\ 
\midrule
CheXCOVID & 3 & Age & 10.8 \\
CheXpert  & 2 & Age & 9.6 \\
Fitzpatrick17k & 3 & Skin & 5.4 \\
HAM10000 & 7 & Age & 58.3 \\
PAPILA & 3 & Age & 5.2 \\
OL3I & 2 & Age & 22.1 \\
\bottomrule
\end{tabular}
\label{table:datasets}
\end{table}

\input{SupplementaryFile/PAPILA_analysis}

\input{SupplementaryFile/holdout_bars_AUC}

\newcolumntype{M}[1]{>{\centering\arraybackslash}m{#1}}

\begin{table*}[h!]
\caption{\textbf{Samples of Datasets.} Samples from the six datasets used in our study show the diversity and real-world relevance of the selected medical imaging tasks.}
\centering
\setlength{\tabcolsep}{10pt} 
\renewcommand{\arraystretch}{2} 
\begin{tabular}{|M{2.4cm}|M{2.3cm}|M{2.3cm}|M{2.3cm}|}

\hline
\textbf{CheXCOVID} & 
\vspace{3mm}\includegraphics[width=0.18\textwidth]{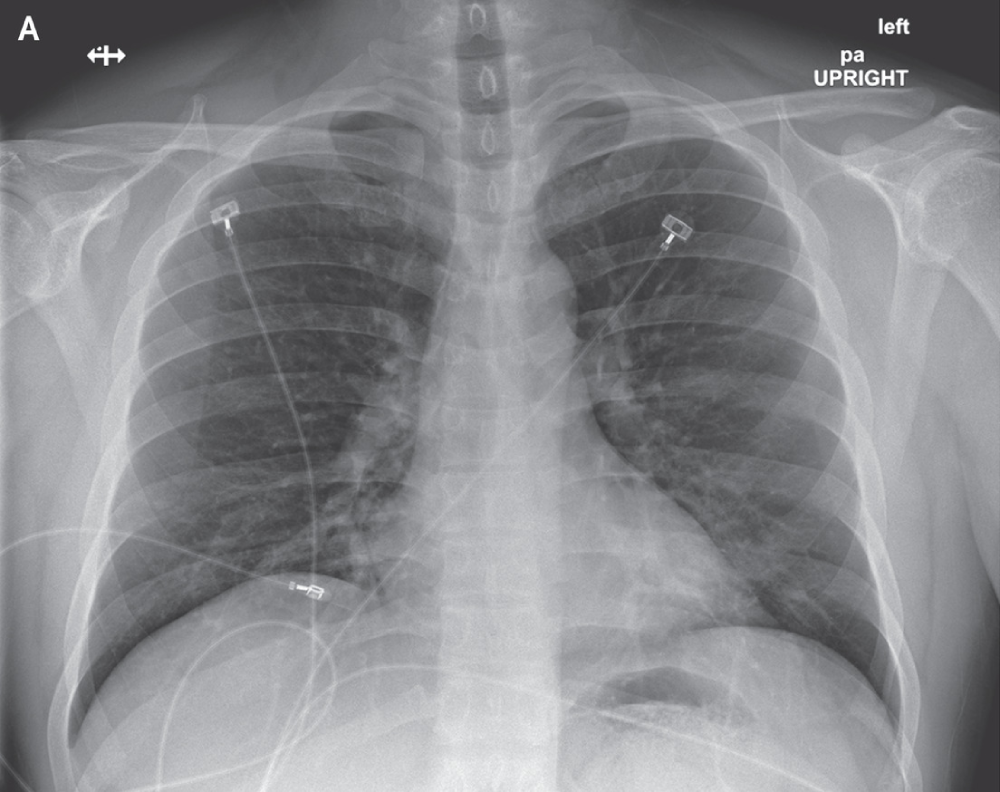} &
\vspace{3mm}\includegraphics[width=0.18\textwidth]{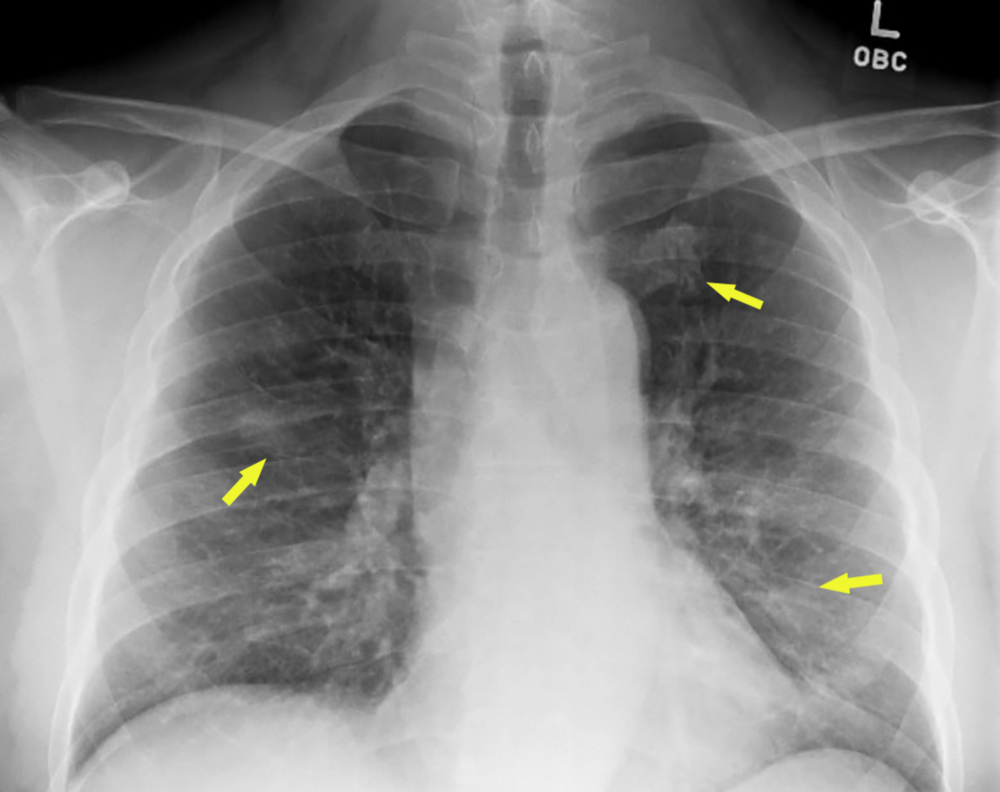} &
\vspace{3mm}\includegraphics[width=0.18\textwidth]{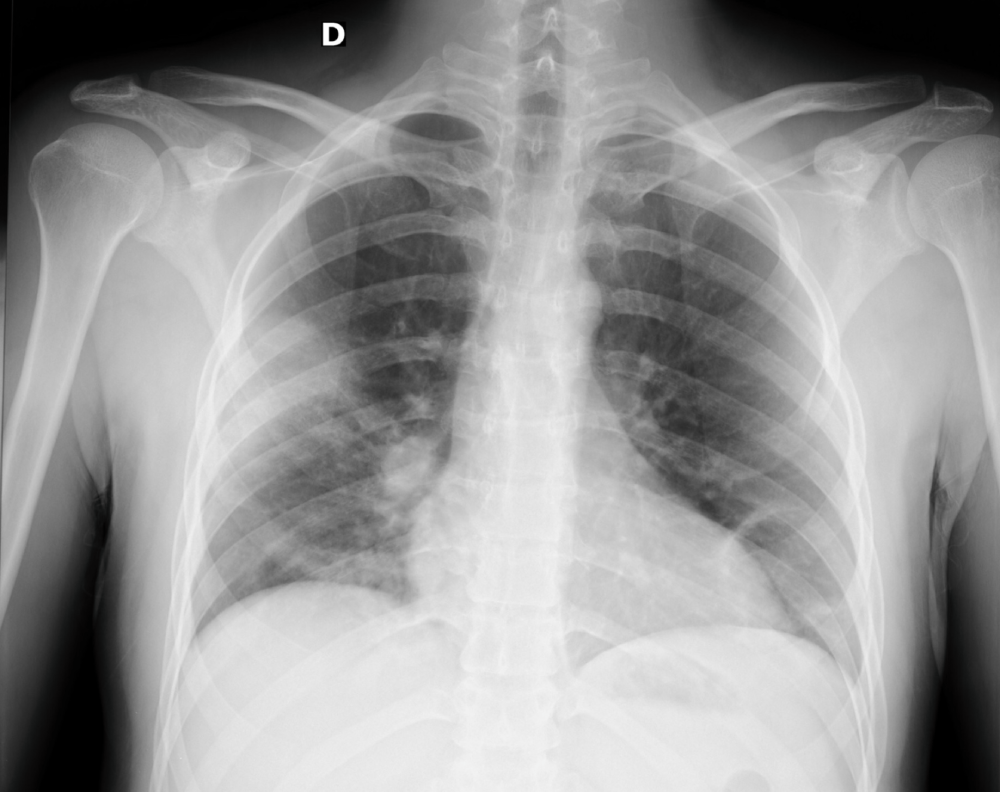} \\ 
\hline
\textbf{CheXpert} & 
\vspace{3mm}\includegraphics[width=0.18\textwidth]{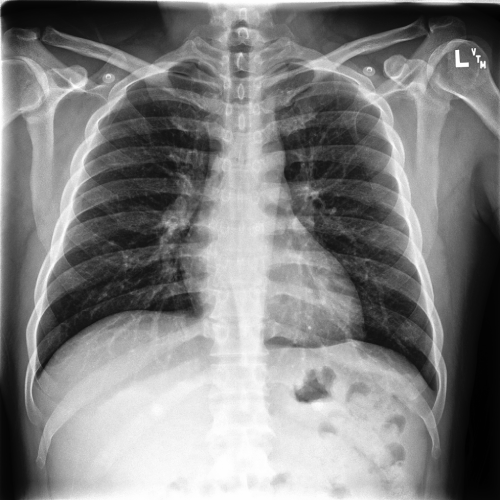} & 
\vspace{3mm}\includegraphics[width=0.18\textwidth]{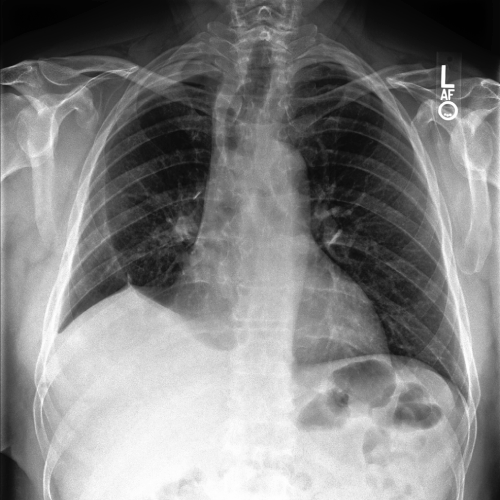} & 
\vspace{3mm}\includegraphics[width=0.18\textwidth]{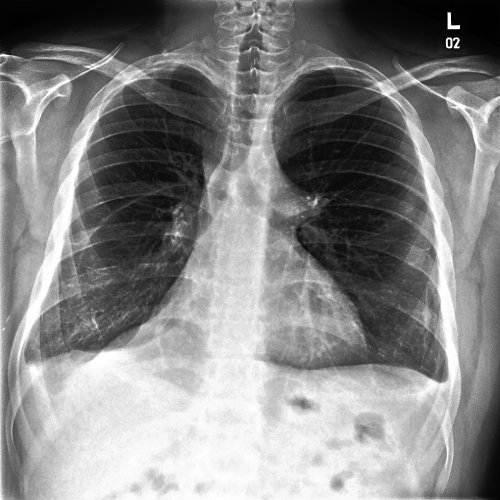} \\ 
\hline
\textbf{Fitzpatrick17k} & 
\vspace{3mm}\includegraphics[width=0.18\textwidth]{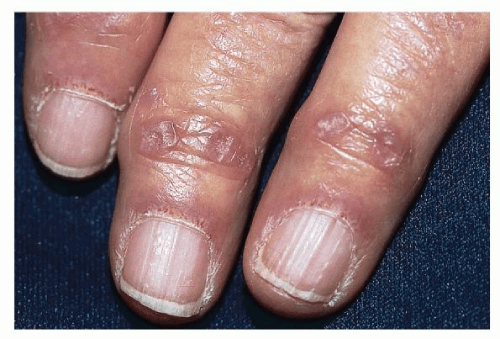} & 
\vspace{3mm}\includegraphics[width=0.18\textwidth]{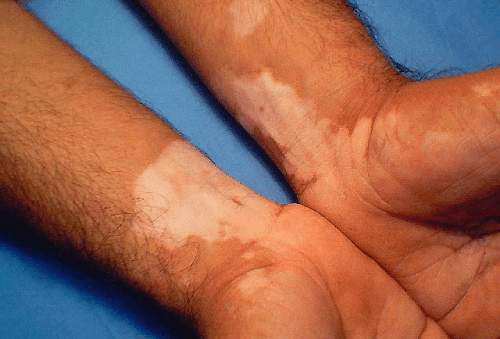} & 
\vspace{3mm}\includegraphics[width=0.18\textwidth]{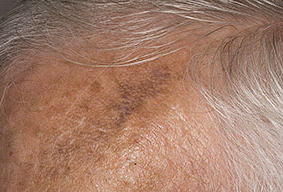} \\ 
\hline
\textbf{HAM10000} & 
\vspace{3mm}\includegraphics[width=0.18\textwidth]{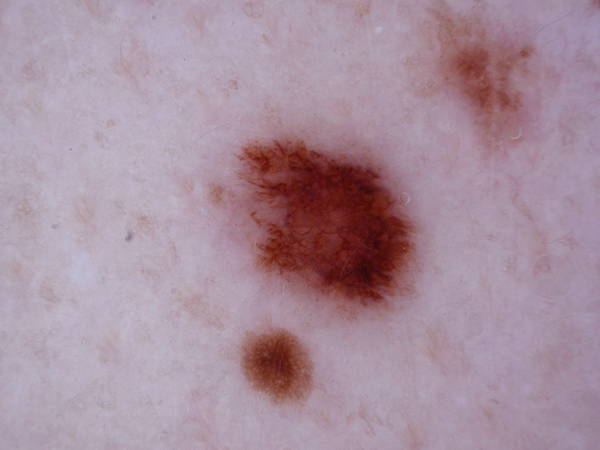} & 
\vspace{3mm}\includegraphics[width=0.18\textwidth]{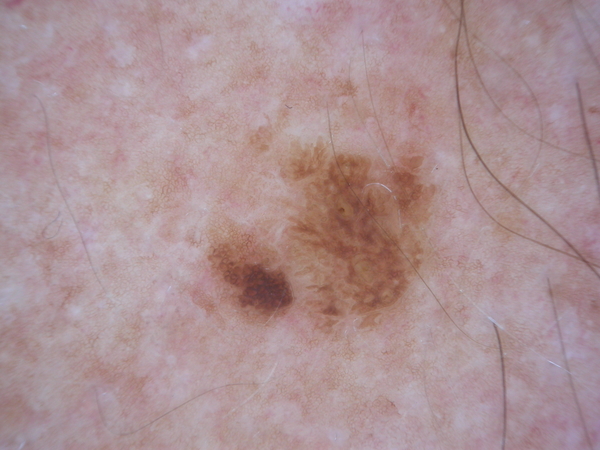} & 
\vspace{3mm}\includegraphics[width=0.18\textwidth]{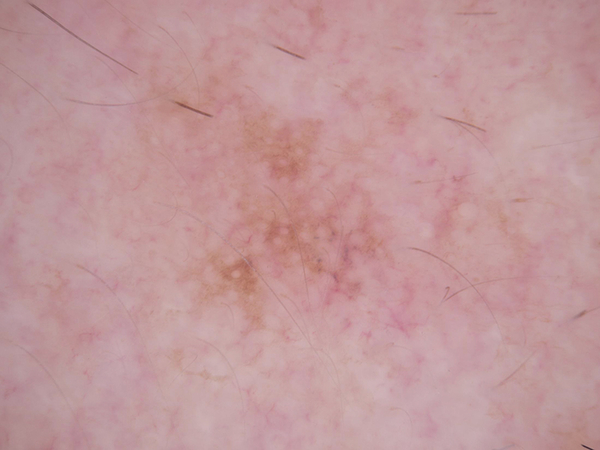} \\ 
\hline
\textbf{PAPILA} & 
\vspace{3mm}\includegraphics[width=0.18\textwidth]{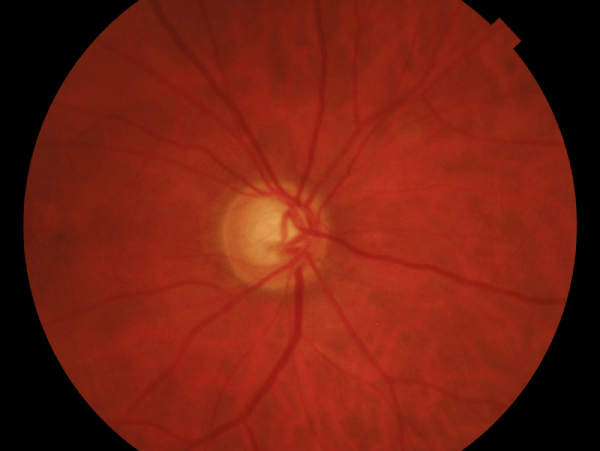} & 
\vspace{3mm}\includegraphics[width=0.18\textwidth]{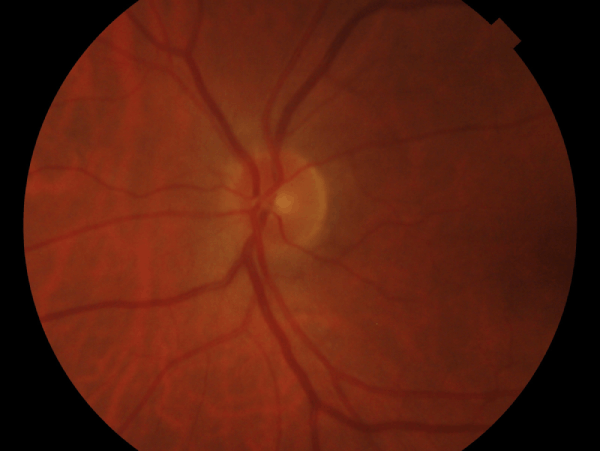} & 
\vspace{3mm}\includegraphics[width=0.18\textwidth]{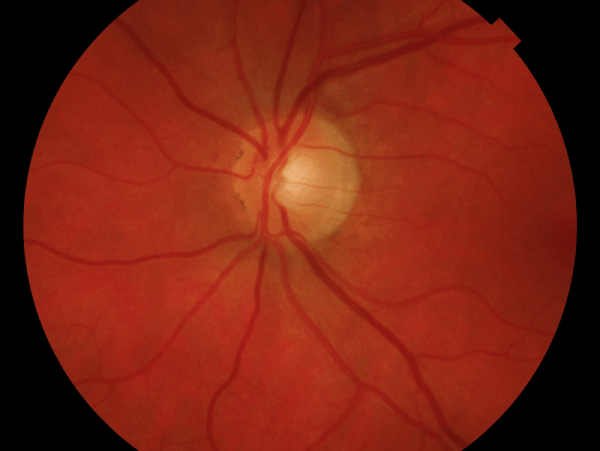} \\ 
\hline
\textbf{OL3I} & 
\vspace{3mm}\includegraphics[width=0.18\textwidth]{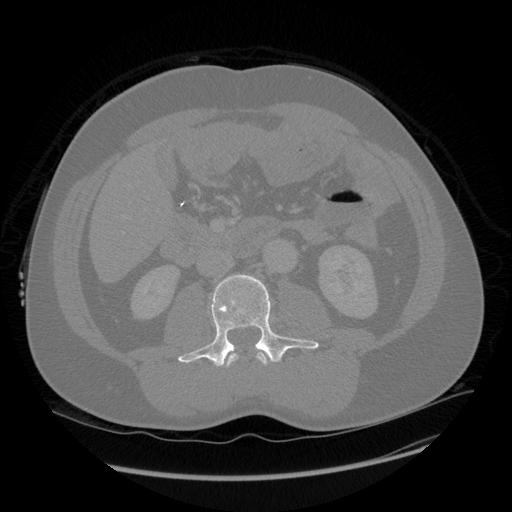} & 
\vspace{3mm}\includegraphics[width=0.18\textwidth]{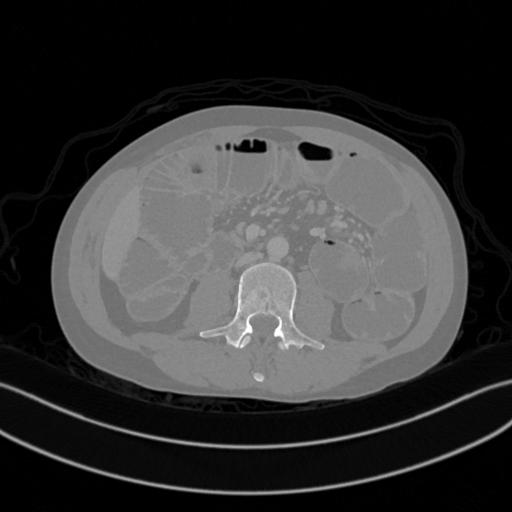} & 
\vspace{3mm}\includegraphics[width=0.18\textwidth]{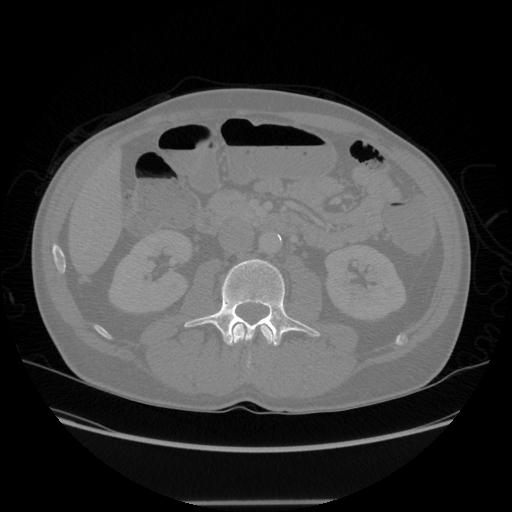} \\ 
\hline
\end{tabular}
\label{table:datasets_visualization}
\end{table*}

%% file: SupplementaryFile/PAPILA_analysis.tex
\begin{figure}[ht]
    \centering
    \includegraphics[width=\textwidth]{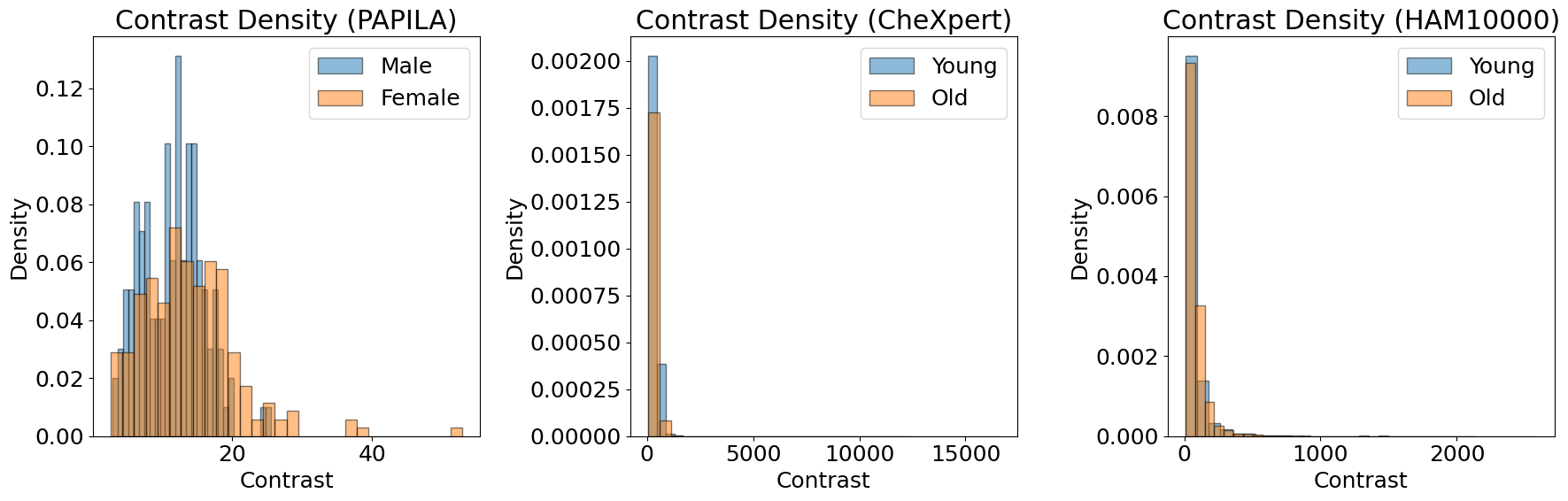}
    \caption{Comparing pixel contrast density different datasets, we observe the significant difference in image characteristics across age groups in PAPILA.}
    \label{fig:PAPILA_analysis}
\end{figure}

%% file: SupplementaryFile/holdout_bars_AUC.tex
\begin{figure}[!ht]   
    \centering
    \includegraphics[width=\textwidth]{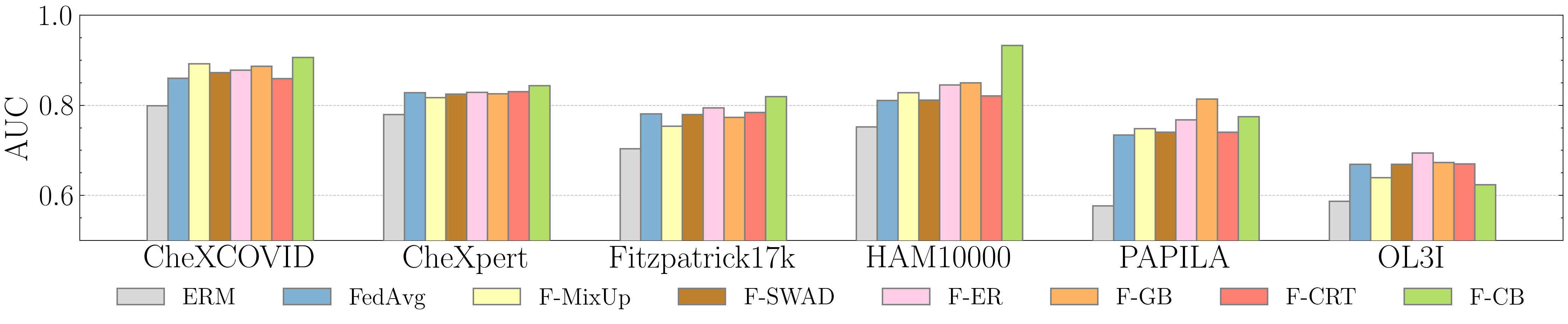}
    \caption{\textbf{AUC Evaluation on New Demographic Distributions.} 
    Reporting LTR (see \Cref{fig:ood_group_incremental}) is important because a model trained on a severely imbalanced dataset (e.g. ERM on HAM10000) can learn to over-predict the majority class, resulting in an AUC (this figure) value much higher than the LTR value.
    }
      \label{fig:ood_auc}
\end{figure}

%% file: Paper.bbl
\begin{thebibliography}{10}
\providecommand{\url}[1]{\texttt{#1}}
\providecommand{\urlprefix}{URL }
\providecommand{\doi}[1]{https://doi.org/#1}

\bibitem{alshammari2022long}
Alshammari, S., Wang, Y.X., Ramanan, D., Kong, S.: Long-tailed recognition via weight balancing. In: Conference on Computer Vision and Pattern Recognition (2022)

\bibitem{baghestani2005skin}
Baghestani, S., Zare, S., Mahboobi, A.A.: Skin disease patterns in hormozgan, iran. International journal of dermatology  \textbf{44}(8),  641--645 (2005)

\bibitem{cavallotti2008age}
Cavallotti, C., Cerulli, L.: Age-related changes of the human eye. Springer Science \& Business Media (2008)

\bibitem{cha2021swad}
Cha, J., Chun, S., Lee, K., Cho, H.C., Park, S., Lee, Y., Park, S.: Swad: Domain generalization by seeking flat minima. Advances in Neural Information Processing Systems  \textbf{34},  22405--22418 (2021)

\bibitem{chaaban2017demographic}
Chaaban, M.R., Zhang, D., Resto, V., Goodwin, J.S.: Demographic, seasonal, and geographic differences in emergency department visits for epistaxis. Otolaryngology--Head and Neck Surgery  \textbf{156}(1),  81--86 (2017)

\bibitem{chaudhry2019tiny}
Chaudhry, A., Rohrbach, M., Elhoseiny, M., Ajanthan, T., Dokania, P.K., Torr, P.H.S., Ranzato, M.: On tiny episodic memories in continual learning (2019)

\bibitem{cohen2013statistical}
Cohen, J.: Statistical power analysis for the behavioral sciences. Academic press (2013)

\bibitem{cohen2020covid}
Cohen, J.P., Morrison, P., Dao, L., Roth, K., Duong, T.Q., Ghassemi, M.: Covid-19 image data collection: Prospective predictions are the future. arXiv preprint arXiv:2006.11988  (2020)

\bibitem{derakhshani2022lifelonger}
Derakhshani, M.M., Najdenkoska, I., van Sonsbeek, T., Zhen, X., Mahapatra, D., Worring, M., Snoek, C.G.: Lifelonger: A benchmark for continual disease classification. In: International Conference on Medical Image Computing and Computer-Assisted Intervention. pp. 314--324. Springer (2022)

\bibitem{diaz2018don}
D{\'\i}az-Rodr{\'\i}guez, N., Lomonaco, V., Filliat, D., Maltoni, D.: Don't forget, there is more than forgetting: new metrics for continual learning. In: Workshop on Continual Learning, Neural Information Processing Systems (2018)

\bibitem{ebrahimian2022fda}
Ebrahimian, S., Kalra, M.K., Agarwal, S., Bizzo, B.C., Elkholy, M., Wald, C., Allen, B., Dreyer, K.J.: Fda-regulated ai algorithms: trends, strengths, and gaps of validation studies. Academic radiology  (2022)

\bibitem{groh2021evaluating}
Groh, M., Harris, C., Soenksen, L., Lau, F., Han, R., Kim, A., Koochek, A., Badri, O.: Evaluating deep neural networks trained on clinical images in dermatology with the fitzpatrick 17k dataset. In: Conference on Computer Vision and Pattern Recognition. pp. 1820--1828 (2021)

\bibitem{haralick1973textural}
Haralick, R.M., Shanmugam, K., Dinstein, I.H.: Textural features for image classification. IEEE Transactions on systems, man, and cybernetics (6),  610--621 (1973)

\bibitem{irvin2019chexpert}
Irvin, J., Rajpurkar, P., Ko, M., Yu, Y., Ciurea-Ilcus, S., Chute, C., Marklund, H., Haghgoo, B., Ball, R., Shpanskaya, K., et~al.: Chexpert: A large chest radiograph dataset with uncertainty labels and expert comparison. In: Proceedings of the AAAI conference on artificial intelligence (2019)

\bibitem{kamnitsas2017efficient}
Kamnitsas, K., Ledig, C., Newcombe, V.F., Simpson, J.P., Kane, A.D., Menon, D.K., Rueckert, D., Glocker, B.: Efficient multi-scale 3d cnn with fully connected crf for accurate brain lesion segmentation. Medical image analysis  \textbf{36},  61--78 (2017)

\bibitem{kang2019decoupling}
Kang, B., Xie, S., Rohrbach, M., Yan, Z., Gordo, A., Feng, J., Kalantidis, Y.: Decoupling representation and classifier for long-tailed recognition. In: International Conference on Learning Representations (2019)

\bibitem{kovalyk2022papila}
Kovalyk, O., Morales-S{\'a}nchez, J., Verd{\'u}-Monedero, R., Sell{\'e}s-Navarro, I., Palaz{\'o}n-Cabanes, A., Sancho-G{\'o}mez, J.L.: Papila: Dataset with fundus images and clinical data of both eyes of the same patient for glaucoma assessment. Scientific Data  \textbf{9}(1), ~291 (2022)

\bibitem{marrakchi2021fighting}
Marrakchi, Y., Makansi, O., Brox, T.: Fighting class imbalance with contrastive learning. In: International Conference on Medical Image Computing and Computer-Assisted Intervention. Springer (2021)

\bibitem{mcmahan2017communication}
McMahan, B., Moore, E., Ramage, D., Hampson, S., y~Arcas, B.A.: Communication-efficient learning of deep networks from decentralized data. In: Artificial intelligence and statistics. pp. 1273--1282. PMLR (2017)

\bibitem{paek2012skin}
Paek, S.Y., Koriakos, A., Saxton-Daniels, S., Pandya, A.G.: Skin diseases in rural yucatan, mexico. International journal of dermatology  \textbf{51}(7),  823--828 (2012)

\bibitem{pekmezaris2013aging}
Pekmezaris, R., Kozikowski, A., Moise, G., Clement, P.A., Hirsch, J., Kraut, J., Levy, L.C.: Aging in suburbia: An assessment of senior needs. Educational Gerontology  \textbf{39}(5),  355--365 (2013)

\bibitem{petkova2020pooling}
Petkova, E., Antman, E.M., Troxel, A.B.: Pooling data from individual clinical trials in the covid-19 era. Jama  \textbf{324}(6),  543--545 (2020)

\bibitem{pooch2020can}
Pooch, E.H., Ballester, P., Barros, R.C.: Can we trust deep learning based diagnosis? the impact of domain shift in chest radiograph classification. In: Thoracic Image Analysis: Second International Workshop, TIA 2020, Held in Conjunction with MICCAI 2020, Lima, Peru, October 8, 2020, Proceedings 2. Springer (2020)

\bibitem{Sagawa*2020Distributionally}
Sagawa*, S., Koh*, P.W., Hashimoto, T.B., Liang, P.: Distributionally robust neural networks. In: International Conference on Learning Representations (2020)

\bibitem{sun2020impacts}
Sun, Z., Zhang, H., Yang, Y., Wan, H., Wang, Y.: Impacts of geographic factors and population density on the covid-19 spreading under the lockdown policies of china. Science of The Total Environment  \textbf{746},  141347 (2020)

\bibitem{ogier2022flamby}
Ogier~du Terrail, J., Ayed, S.S., Cyffers, E., Grimberg, F., He, C., Loeb, R., Mangold, P., Marchand, T., Marfoq, O., Mushtaq, E., et~al.: Flamby: Datasets and benchmarks for cross-silo federated learning in realistic healthcare settings. Advances in Neural Information Processing Systems  (2022)

\bibitem{tschandl2018ham10000}
Tschandl, P., Rosendahl, C., Kittler, H.: The ham10000 dataset, a large collection of multi-source dermatoscopic images of common pigmented skin lesions. Scientific data  (2018)

\bibitem{vokinger2021continual}
Vokinger, K.N., Feuerriegel, S., Kesselheim, A.S.: Continual learning in medical devices: Fda's action plan and beyond. The Lancet Digital Health  (2021)

\bibitem{white2022data}
White, T., Blok, E., Calhoun, V.D.: Data sharing and privacy issues in neuroimaging research: Opportunities, obstacles, challenges, and monsters under the bed. Human Brain Mapping  (2022)

\bibitem{yala2019deep}
Yala, A., Lehman, C., Schuster, T., Portnoi, T., Barzilay, R.: A deep learning mammography-based model for improved breast cancer risk prediction. Radiology  \textbf{292}(1),  60--66 (2019)

\bibitem{yang2023change}
Yang, Y., Zhang, H., Katabi, D., Ghassemi, M.: Change is hard: A closer look at subpopulation shift. In: International Conference on Machine Learning (2023)

\bibitem{zambrano2021opportunistic}
Zambrano~Chaves, J.M., Chaudhari, A.S., Wentland, A.L., Desai, A.D., Banerjee, I., Boutin, R.D., Maron, D.J., Rodriguez, F., Sandhu, A.T., Jeffrey, R.B., et~al.: Opportunistic assessment of ischemic heart disease risk using abdominopelvic computed tomography and medical record data: a multimodal explainable artificial intelligence approach. medRxiv pp. 2021--01 (2021)

\bibitem{zhang2018mixup}
Zhang, H., Cisse, M., Dauphin, Y.N., Lopez-Paz, D.: mixup: Beyond empirical risk minimization. In: International Conference on Learning Representations (2018)

\bibitem{zong2023medfair}
Zong, Y., Yang, Y., Hospedales, T.: Medfair: Benchmarking fairness for medical imaging. In: International Conference on Learning Representations (2023)

\end{thebibliography}
